\documentclass[doublecol]{epl2} 
\usepackage{dsfont}
\usepackage{epsf}
\usepackage{epstopdf}
\usepackage{hyperref}
\usepackage[latin1]{inputenc}
\usepackage{amssymb,amsmath}

\usepackage{setspace}
\usepackage{sidecap}
\usepackage{amsfonts}
\usepackage{dsfont}
\usepackage{mathtools}
\usepackage{verbatim}
\usepackage{epstopdf}
\usepackage{latexsym}
\usepackage{dcolumn}
\usepackage{epsf}
\usepackage{float}
\DeclarePairedDelimiter\abs{\lvert}{\rvert}

\usepackage{color}
\usepackage[latin1]{inputenc}
\usepackage{graphicx}

\usepackage{comment}
\title{Perfect synchronization in networks of phase-frustrated  oscillators}

\author{Prosenjit Kundu,\inst{1} Chittaranjan Hens,\inst{2}  
Baruch Barzel \inst{2} \and Pinaki Pal\inst{1}}

\shortauthor{Prosenjit Kundu \etal}

\institute{                    
  \inst{1} Department of Mathematics, National Institute of Technology, Durgapur 713209, India\\
  \inst{2} Department of Mathematics, Bar-Ilan University, Ramat-Gan, 52900, Israel
}
\pacs{05.45.Xt}{Synchronization; coupled oscillators}
\pacs{89.75.-k}{Complex systems}
 	 
\abstract{
Synchronizing phase frustrated Kuramoto oscillators, a challenge that has found applications from neuronal networks to the power grid, is an eluding problem, as even small phase-lags cause the oscillators to avoid synchronization. Here we show, constructively, how to strategically select the optimal frequency set, capturing the natural frequencies of all oscillators, for a given network and phase-lags, that will ensure perfect synchronization. We find that high levels of synchronization are sustained in the vicinity of the optimal set, allowing for some level of deviation in the frequencies without significant degradation of synchronization. Demonstrating our results on first and second order phase-frustrated Kuramoto dynamics, we implement them on both model and real power grid networks, showing how to achieve synchronization in a phase frustrated environment.
}
\begin{document}

\maketitle

Synchronization captures the emergence of collective behavior in complex systems \cite{Arenas-physrep,Pikovsky,Boccaletti-physrep}, ranging from neuronal dynamics \cite{Belykh-Burst} to animal behavior \cite{Strogatz 2003} and technological networks \cite{Motter}. In its classic formulation synchronization is driven by the coupling between the oscillators, which drives them towards collective oscillations, overcoming the diversity in the intrinsic frequencies of each individual oscillator \cite{Kuramoto,Bonilla,Arenas,Moreno,Skardal}. Hence synchronization is enhanced either by increasing the coupling strength between the oscillating units, or by a homogeneous frequency distribution among all oscillators. These strategies towards synchronization, however, fail in case the coupling between the oscillators induces phase-lags \cite{Sakaguchi,Omel'chenko,Kurths-Physreport}, a common characteristic featured by many real systems, where the components take time to respond to their neighboring oscillators. Indeed, under phase-frustration, the system persistently avoids synchronization, even when the frequencies are homogeneous or under relatively strong coupling. 

To overcome this challenge, we derive here the link between the network characteristics, the phase-lags and the optimal frequency set, that allows the phase-frustrated system to reach perfect synchronization. This allows us a two way prediction: for a given network and phase-lags, we predict the optimal selection of natural frequencies that will ensure synchronization. Alternatively, given a set of natural frequencies - we show how to design the network that will lead the oscillators towards perfect synchrony. We find, numerically, that our predicted synchronization is quite robust, exhibiting a range of phase-locked solutions even under deviations from our predicted frequencies/networks, thus being insensitive to moderate levels of noise/perturbation. Counter-intuitively, we find that synchronization is not necessarily enhanced by strengthening the coupling or by selecting homogeneous frequencies, rather it emerges from the complex interplay between the selected frequencies, the distribution of phase-lags and the structure of the weighted underlying network. 

Consider  a system of $N$ coupled oscillators, whose phases $\theta_i(t)$, ($i = 1\dots N$), are driven by the dynamic equations \cite{Kuramoto,Sakaguchi}
\begin{eqnarray}\label{eqn1}
\frac{d\theta_i}{dt} &=& \omega_i +  \sum_{j=1}^{N} A_{ij}{F}(\theta_j - \theta_i-\alpha_{ij}), 
\end{eqnarray}
\noindent
where $\omega_i$ represents node $i$'s natural frequency and $A_{ij}$ is a weighted adjacency matrix with arbitrary degree and weight distributions. The functional form of the coupling is captured by $F(\theta_j - \theta_i - \alpha_{ij})$, a $2 \pi$ periodic function, with distributed phase-lags $\alpha_{ij}$, which capture the response time of oscillator $i$ to changes in its neighbor's phase $\theta_j$. Phase-frustrated models of the form (\ref{eqn1}) are frequently used to describe coupled systems, from Josephson junctions \cite{Wissenfeld} to power supply networks \cite{Dorfler-siam} and mechanical rotors \cite{David-rotor}. Choosing $F(\theta) = \sin(\theta - \alpha)$ (with $\alpha_{ij}=\alpha$ independent of $i$ and $j$), Eq.\ (\ref{eqn1}) converges to the Sakaguchi-Kuramoto model of phase frustrated oscillators \cite{Sakaguchi,Omel'chenko,Kurths-Physreport,Lohe}; setting $\alpha \rightarrow 0$ we arrive at the classic Kuramoto dynamics \cite{Kuramoto}.
 
\par To quantify the level of synchronization in the system we use the Kuramoto order parameter 
\begin{eqnarray}
r = \left| \frac{1}{N}\sum_{j=1}^{N} e^{i\theta_j} \right|,
\label{r}
\end{eqnarray}
which approaches $r = 0$ in the disordered regime and $r = 1$ in the limit where all oscillators are in perfect synchrony, namely $\theta_1 = \theta_2 = \dots = \theta_N$. In the classic Kuramoto framework the level of synchronization is determined by the trade-off between the heterogeneity of the natural frequencies $\omega_i$ and the strength of the coupling, as determined by $A_{ij}$. Hence to achieve synchronous behavior one draws $\omega_i$ from a narrowly bounded distribution ({\it e.g.}, normal distribution) or increases the weights of $A_{ij}$ until reaching $r \rightarrow 1$. Such perfect synchronization, however, is unattainable in the presence of phase-lags even for extreme coupling strengths \cite{Skardal-Pre14, Skardal-PhysicaD}. Hence we seek the optimal frequency sequence ${\boldsymbol \omega} = (\omega_1, \omega_2,....,\omega_N)^\top $ that will enable perfect synchronization $r = 1$ for phase-frustrated oscillators of the form (\ref{eqn1}).
 
\begin{figure}
\includegraphics[height=4.0cm,width=8.5cm]{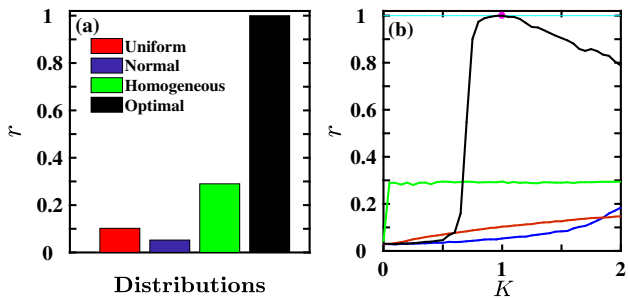}
\caption{{\bf Synchronizing phase-frustrated oscillators}. 
(a) We tested the synchronization $r$ of phase-frustrated scale-free oscillator network with different selections of $\boldsymbol{\omega}$ by numrically solving Eq.\ (\ref{eqn1}). The classic choices, uniformly distributed (red), normally distributed (blue) or homogeneous (green) cannot lead to synchronization. The optimal $\boldsymbol{\omega}$ from (\ref{wt_freq_set2}), however, leads to global synchronization with $r = 1$, as predicted. 
(b) Increasing the coupling strengths $A_{ij}$ by a factor $K$ we find that the system persistently avoids synchronization even for large $K$, under the general frequency distributions (red, blue, green). Synchronization is only obtained using our optimal $\boldsymbol{\omega}$ (black) as predicted by (\ref{wt_freq_set2}). Eq.\ (\ref{wt_freq_set2}) also predicts that any change in $A_{ij}$ will harm synchronization, hence perfect synchronization ($r = 1$, magenta dot) is only obtained for ($K = 1$). However our optimal $\boldsymbol{\omega}$ allows for a phase-locked solution ($r \sim 1$) for broad values of $K$ around $K = 1$. The onset of the phase-locked solutions occurs rather sharply at $K \approx 0.78$ (vertical yellow line), as predicted below in Fig.\ \ref{stability}b. 
}	
\label{fig:r vs k}
\end{figure}

To obtain ${\boldsymbol \omega}$, we analyze Eq.\ (\ref{eqn1}) as it approaches synchronization \cite{Skardal, Skardal-Pre14, Skardal-PhysicaD, Brede}, namely in the limit where $|\theta_j - \theta_i| \rightarrow 0$. In this limit, the coupling function in (\ref{eqn1}) can be approximated by $F(\theta_j - \theta_i - \alpha_{ij}) \approx F(-\alpha_{ij}) + F^{\prime}(-\alpha_{ij})(\theta_j - \theta_i)$, where $F^{\prime}(\alpha) = \left. \frac{\partial F}{\partial\theta} \right|_{\theta = 0}$. This allows us to write Eq.\ (\ref{eqn1}) as
\begin{eqnarray}
\frac{d\theta_i}{dt} & = & 
\omega_i + d_{i}- \sum_{j=1}^{N}L_{ij}\theta_j,
\label{eqn3}
\end{eqnarray}
where 
\begin{eqnarray}
{{d}_i} = \sum_{j=1}^{N}A_{ij} F(-\alpha_{ij})
\label{degree}
\end{eqnarray}
and
\begin{eqnarray}
{L}_{ij}=\delta_{ij}\bigg(\sum_{q=1}^{N} A_{iq}F^{\prime}(-\alpha_{iq})\bigg)- A_{ij} F^{\prime}(-\alpha_{ij}),
\label{eqn3b}
\end{eqnarray}
in which $\delta_{ij}$ is the Kronecker $\delta$-function. The system will reach a synchronized state if, for some choice of the natural frequencies ${\boldsymbol \omega}$, Eq.\ (\ref{eqn3}) reaches a stable solution in which all phases, $\theta_i(t)$, evolve according to some common frequency $\Omega$, namely $\theta_i(t) = \phi_{i}+\Omega t$, where $\phi_{i}=\theta_i(0)$ is the phase of oscillator $i$. Transforming to an $\Omega$-rotating frame, we have $\frac{d\theta_i}{dt}=0$, which in Eq.\ (\ref{eqn3}) leads to
\begin{equation}
{\boldsymbol {\phi}} = { L^\dagger}{({\boldsymbol{\omega}} +  {\boldsymbol{d}})},
\label{fp1}
\end{equation} 
\noindent
where ${\boldsymbol {\phi}}=(\phi_{1},\phi_{2},....,\phi_{N})^\top$ represents  the vector of all phases, ${\boldsymbol {d}}=(d_{1},d_{2},....,d_{N})^\top$ and ${{ L}^\dagger}$ is the pseudo-inverse \cite{ana} of ${L}$ in (\ref{eqn3b}). The condition (\ref{fp1}) captures frequency synchronization, a state in which all units oscillate at a common frequency, but with different phases $\phi_i$. Complete synchronization, however, requires also that all phases condense around a common value $\phi$, which, by additional rotation of the system can be set to ${\boldsymbol {\phi}}=0$. We this gauging we arrive at 
\begin{equation}
{\boldsymbol{\omega}} = -{\boldsymbol{{d}}}+ C \mathbf{1},
\label{wt_freq_set}
\end{equation}
where $\mathbf{1}=(1,...,1)^\top$ and $C$ is an arbitrary constant, a degree of freedom enabled due to the fact that $ L^{\dagger} \mathbf{1}=0$ \cite{ana}. We use this degree of freedom to select $C= \langle{d}\rangle$, allowing us to write, explicitly, the designated frequency of the $i$th oscillator as
\begin{equation}
\omega_{i} = -\sum_{j=1}^{N} A_{ij}F(-\alpha_{ij})+\frac{1}{N} \sum_{i,j=1}^{N} A_{ij}F(-\alpha_{ij}),
\label{wt_freq_set2}
\end{equation} 
for which $\langle {\boldsymbol{\omega}}\rangle=\frac{1}{N} \sum_{i=1}^{N} \omega_{i}=0$, namely we gauge the mean frequency to be zero. 
   
\par Equation (\ref{wt_freq_set2}) represents our key prediction, providing the optimal frequency set ${\boldsymbol{\omega}}$ in a weighted network of heterogeneous phase-frustrated oscillators. It indicates that the optimal frequency set $\boldsymbol {\omega}$ is determined by the interplay between the system's topology ($A_{ij}$), its dynamics ($F$) and the specific form of the distributed phase-lags ($\alpha_{ij}$). For the un-frustrated Kuramoto model $(F(\theta)=\sin(\theta),\alpha=0)$ it predicts that the optimal frequency set is uniform, $w_i = 0$ for all $i$, reaffirming Kuramoto's classic prediction \cite{Kuramoto}. In the Sakaguchi-Kuramoto model ($\alpha_{ij}=\alpha$) Eq.\ (\ref{wt_freq_set2}) predicts that $\omega_i$ scales with node $i$'s weighted degree $\omega_i \sim \sum_{j=1}^N A_{ij}$ up to an additive constant $C=\frac{F(-\alpha)}{N}\sum_{i,j=1}^N A_{ij}$. This implies that contrary to the Kuramoto model, where synchrony is a consequence of $\boldsymbol{\omega}$'s homogeneity, in the phase-frustrated case $\boldsymbol{\omega}$  depends on $A_{ij}$'s degree sequence, therefore it must follow $A_{ij}$'s degree heterogeneity. For instance, if $A_{ij}$ is scale-free, as often encountered in real networks \cite{Albert,Cohen}, $\omega_i$ must also be drawn from a scale-free distribution. Hence, counter-intuitively, (\ref{wt_freq_set2}) shows that perfect synchrony may arise from oscillator heterogeneity, namely from a scale-free sequence ${\boldsymbol{\omega}}$.  

\par To test our prediction we constructed Eq.\ (\ref{eqn1}) using a weighted scale-free network $A_{ij}$ ($P(k)\sim k^{-\gamma}, \gamma=3$) of $N = 1,000$ interacting oscillators, whose phase-lags were extracted from a uniform distribution $0.1 \le \alpha_{ij} \le \pi/2$, \textit{i.e.} $\alpha_{ij} \sim \mathcal{U}(0.1,1.57)$. The weights of all existing links were also extracted from a uniform distribution $A_{ij}\sim \mathcal {U}(0.1,1.5)$. We then numerically solved Eq.\ \ref{eqn1}) and tested the level of synchronization $r$ (\ref{r}), for several choices of ${\boldsymbol{\omega}}$: {\it homogeneous}, in which all $\omega_i$ are identical; {\it normal}, in which $\omega_i$ are extracted from a normal distribution with mean $0$ and variance $1$, \textit{i.e.} $\omega_{i}\sim \mathcal{N}(0,1)$; and {\it uniform}, where $\omega_{i}\sim \mathcal{U}(-2,2)$. For uniform ${\boldsymbol{\omega}}$ we find that the system cannot synchronize with $r$ being significantly smaller than unity (Fig.\ \ref{fig:r vs k}(a), red). Synchronization becomes even lower for normal (blue), and slightly improved for homogeneous (green). Hence, as opposed to the un-frustrated Kuramoto dynamics, here a bounded frequency distribution cannot lead to synchronization. Our theory, however, predicts that synchronization can be obtained if we construct ${\boldsymbol{\omega}}$ using the optimal frequency set (\ref{wt_freq_set2}). Indeed, we find that selecting our predicted ${\boldsymbol{\omega}}$, the system successfully reaches perfect synchronization, featuring $r = 1$, as predicted (black). 
   
\par As explained above, synchronization is often a consequence of two competing effects: the strength of the coupling, which forces the system into collective oscillations vs.\ the heterogeneity in ${\boldsymbol{\omega}}$, which drives the system away from synchronization. We now test these two effects systematically, by first, rescaling, and hence weakening/strengthening, the coupling between all oscillators, and then adding increasing levels of noise to the optimal frequency set predicted in (\ref{wt_freq_set2}).

\noindent {\it Coupling strength.} Often one wishes to {\it force} a system towards global synchrony by strengthening the level of coupling between the interacting oscillators, for instance, multiplying all $A_{ij}$ terms (weights) by a factor of $K > 1$. For phase-frustrated systems of the form (\ref{eqn1}), however, such approach will not lead to global synchrony. Indeed, as Fig.\ \ref{fig:r vs k}(b) indicates, for ${\boldsymbol{\omega}}$ uniform (red), normal (blue) and homogeneous (green), the system consistently avoids global synchronization, despite increasing $K$. For the optimal frequency set (black), we obtain perfect synchronization for $K = 1$ (magenta dot), as predicted, yet increasing or decreasing $K$ harms the level of synchronization since any change to $A_{ij}$, even increasing the strength of the coupling ($K > 1$), leads to consequent changes in the optimal frequency set (\ref{wt_freq_set2}). The important point is that for a rather broad range of $K$  values, the system sustains relatively high levels of synchronization, allowing for a phase-locked solution, even if the selected frequency set is not precisely the optimal one predicted by (\ref{wt_freq_set2}). This represents the robustness of our solution, opening a wide window of phase-locked solutions in the vicinity of the optimal selection (\ref{wt_freq_set2}). 
 
\begin{figure}
\begin{center}
\includegraphics[height=7.1cm,width=7.3cm]{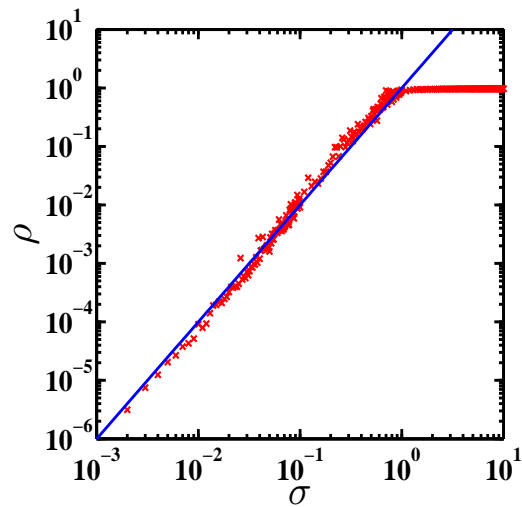}
\caption{{\bf The impact of noise on  synchronization}. We introduce a $\sigma$-deviation from the optimal frequency set, and measure synchronization loss $\rho$ vs.\ $\sigma$ (red). As predicted in Eq.\ (\ref{theo_rho}) we find that in the limit of small $\sigma$, $\rho \sim \sigma^2$ (blue). When $\sigma \gtrsim 1$, the optimal frequency set is overridden by noise, and synchronization is lost completely with $\rho \rightarrow 1$.}
\label{fig:Noise Curve}
\end{center}
\end{figure}
 
\begin{figure}
\includegraphics[height=4.0cm,width=8.0cm]{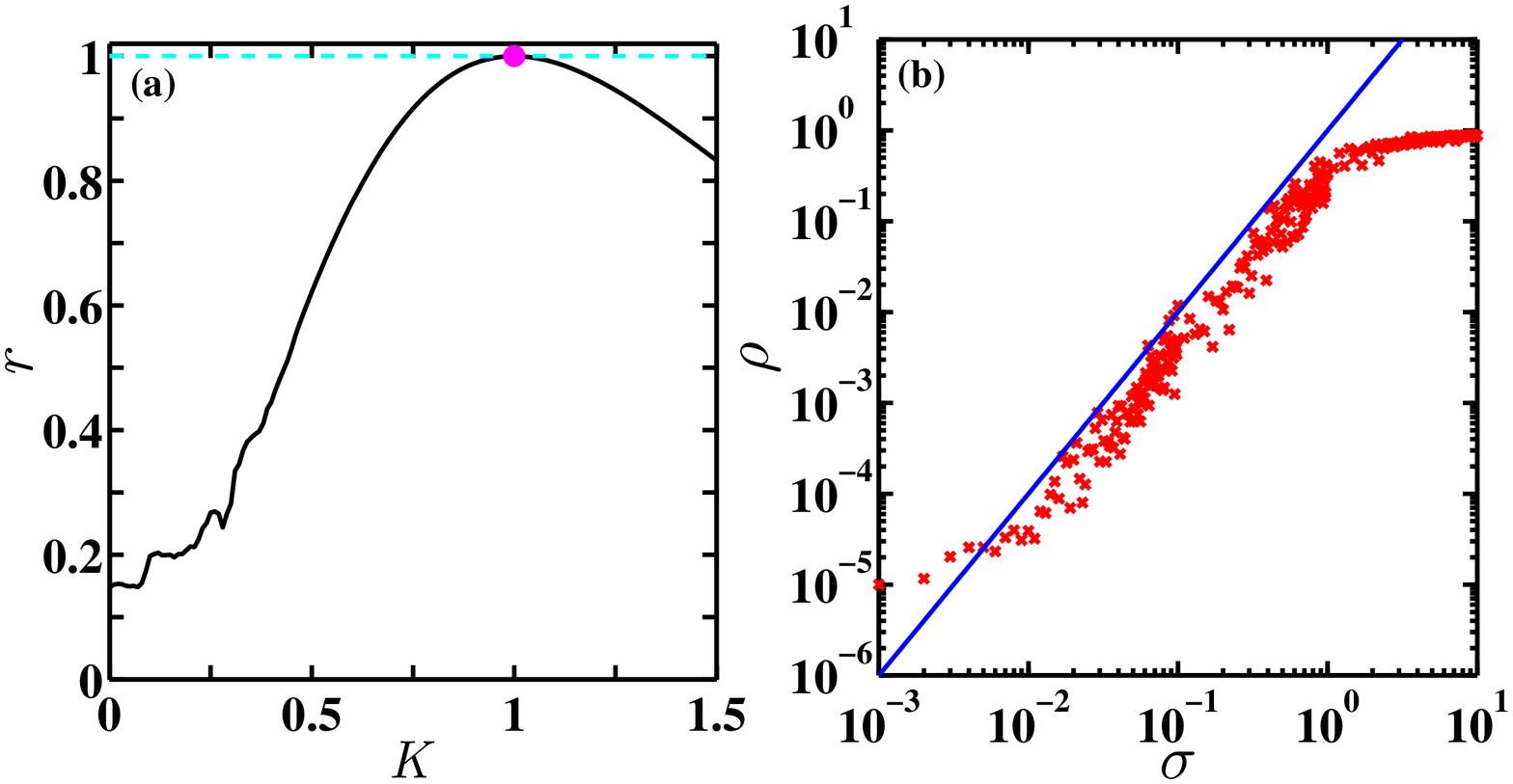}
\includegraphics[height=4.0cm,width=8.0cm]{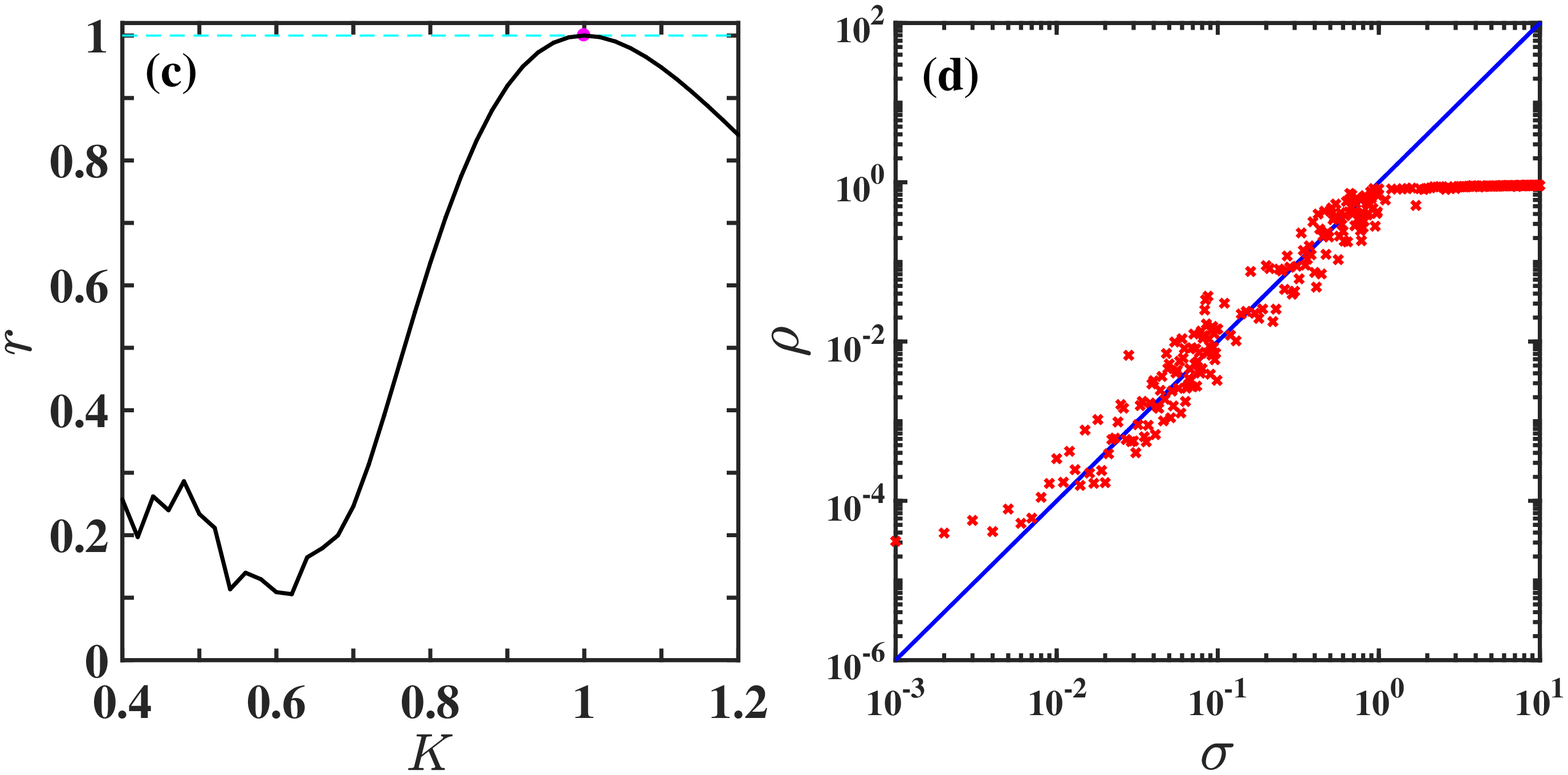}
\caption{{\bf Synchronization in a power supply network}. We constructed the optimal frequency set ${\boldsymbol{\omega}}$ for the Northern European power grid network with second order dynamics of the form (\ref{eqnk21}), setting the damping parameter to $\beta = 10$, \textit{i.e.} strong damping. (a) With this set of frequencies we find a phase-locked solution for a broad range of coupling strengths $K$. The optimal frequencies were constructed for $K = 1$, where  the system reaches perfect synchronization, $r = 1$, as predicted (magenta dot). (b) Adding noise to ${\boldsymbol{\omega}}$ we find that $\rho \sim \sigma^2$, in agreement with (\ref{theo_rho}) and saturates as $\rho \rightarrow 1$ in the limit where  $\sigma \gtrsim 1$.(c),(d) Similar results are also observed under weak damping $\beta=0.1$}.

\label{fig:second order}
\end{figure}
 
\noindent{\it Frequency deviations}. To test the sensitivity of the synchronization to deviations from the optimal frequency set (\ref{wt_freq_set2}) we add Gaussian noise to ${\boldsymbol{\omega}}$, setting ${{\omega_i}} \rightarrow {{\omega_i}} + {\delta{\omega_i}}$, where ${\delta{\omega_i}} \sim \mathcal {N}(0, \sigma{{\omega_i}})$, a random variable extracted from a normal distribution with mean zero and variance ${\sigma^2{\omega_i}^2}$, representing multiplicative noise that is proportional to $\omega_i$. Such deviation will reduce the level of synchronization to $r < 1$, resulting in synchronization loss, which can be quantified by $\rho = 1 - r$ ($0 \le \rho \le 1$). For small $\sigma$ the deviation from synchronization is small, allowing us to approximate (\ref{r}) up to second order as $r \approx 1 - {\lVert{\boldsymbol{\phi}}\rVert}^2/{2N}\nonumber$ \cite{Skardal,Skardal-Pre14}, therefore,  
\begin{eqnarray}
\rho = 1 - r &\sim& \frac{1}{2} \rm {Var} (\phi)
\label{r_SAF}
\end{eqnarray}
where Var$(X)$ represents the variance of the random variable $X$. Using (\ref{fp1}), we write 
\begin{equation}
{\phi_i} = { \sum_j L_{ij}^\dagger} {({\omega_j} +  {{d_j}})} = \sum_j L_{ij}^\dagger\delta{\omega_j}, \label{noise}
\end{equation} 
and hence, with $L_{ij}^\dagger$ being approximately independent of $\delta{\omega_j}$, we have
\begin{equation}
{\rm{Var}} (\phi_{i})= \sum_j (L_{ij}^\dagger)^{2}   {\rm{Var}}(\delta{\omega_{j}}), 
\label{noise2}
\end{equation}
where we also used the fact that $\delta \omega_j$ are independent of each other. As a result we find that
\begin{equation}
{\rm{Var}}(\phi_{i}) \approx C_{i} \sigma^2,
\end{equation}
where $C_{i}=\sum_{j}(L_{ij}^\dagger)^2 \omega_{j}^2$, and hence the overall variance of all $\phi_i$   satisfies $ {\rm{Var}}(\phi) \propto \sigma^2$, where the proportion constant is a function of all pre-factors $C_i$. Omitting such factors that do not depend on the noise level $\sigma$, we arrive at the scaling relationship ${\rm Var(\phi)}\sim \sigma^2$, which in (\ref{r_SAF}) provides
\begin{eqnarray} 
\rho&\sim&\sigma^2,
\label{theo_rho}
\end{eqnarray}  
showing that synchronization loss is quadratically dependent on the noise level in the oscillator frequencies. This allows us to evaluate the decay in synchronization as we deviate from the optimal frequency set (\ref{wt_freq_set2}). For small $\sigma$ we predict the scaling (\ref{theo_rho}) and as $\sigma$ increases $\boldsymbol{\omega}$ continues to deviate from (\ref{wt_freq_set2}) until eventually synchronization is completely lost and $\rho\rightarrow 1$. This behavior is clearly observed in Fig.\ \ref{fig:Noise Curve}, where we introduce increasing levels of noise to the optimal frequency set. We find that for small noise levels $\rho\sim\sigma^2$ (solid line) as predicted and for $\sigma \rightarrow 1$, a limit where $\boldsymbol{\omega}$ is completely overridden by noise, synchronization is lost with $r \rightarrow 0$ and, consequently, $\rho$ approaching unity.

\par 
\noindent
\textit{Second order dynamics}. Our formalism is also applicable beyond the limits of Eq.\ (\ref{eqn1}). To show this we focus on second order phase-frustrated Kuramoto dynamics, captured by
\begin{eqnarray}
\frac{d^2\theta_i}{dt^2} &=& P_i -\beta\frac{d\theta_i}{dt} + \sum_{j=1}^{\mathrm{N}}A_{ij}\mathrm{F}(\theta_j - \theta_i-\alpha_{ij}), 
\label{eqnk21} 
\end{eqnarray}
as frequently used to describe phase synchronization in power supply networks \cite{marc-timme12,Filatrella,marc-timme16,Motter,motter15}. In (\ref{eqnk21}) $P_i$ represents the generated ($P_i > 0$) or consumed ($P_i < 0$) power, and $\beta$ is the damping coefficient of the system components. To examine our formalism in an empirical setting we collected data from the Northern European power grid \cite{Menck}, comprising $N = 236$ nodes and $E = 320$ links. We extracted the frustration terms from a uniform distribution, $\alpha_{ij} \sim \mathcal{U}(0,0.5)$.  As before, we constructed the optimal frequency set ${\boldsymbol{\omega}}$ and tested the level of synchronization $r$ against varying levels of coupling $K$ and noise $\sigma$, setting $\beta = 10$ and $\beta = 0.1$, to represent the limits of strong and weak damping. We find that also in the case of second order dynamics, the empirical power grid network reaches global synchronization ($r = 1$) for $K = 1$, as predicted. It exhibits a phase-locked solution, $r \lesssim 1$, at a range of $K$ values around $K = 1$ (Fig.\ \ref{fig:second order}(a,c)). The loss of synchronization scales as $\rho \sim \sigma^2$, for small deviations from the optimal ${\boldsymbol{\omega}}$, with complete synchronization loss ($\rho \rightarrow 1$) at $\sigma \approx 1$, the point where noise levels become comparable to the frequencies themselves (Fig.\ \ref{fig:second order}(b,d)).

\begin{figure}    
\includegraphics[height=2.9cm,width=8.8cm]{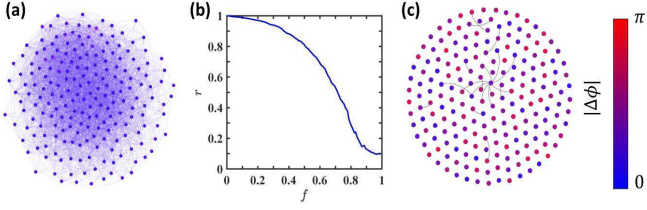}
\caption{{\bf Designing a synchronizable network}.
(a) Starting from a pre-selected frequency set $\boldsymbol{\omega}$, in which $\omega_i \sim \mathcal{U}(2,30)$, we used (\ref{wt_freq_set2}) to construct an optimal network for synchronization. As predicted, this optimal network provides precisely $r = 1$, capturing perfect synchronous oscillations, denoted here by the uniform color of all nodes (blue), which represents $|\Delta \phi_i| = 0$ ($\Delta \phi_i = \phi_i - \langle \phi \rangle$ is the deviation of node $i$'s phase from the mean phase over all nodes, ranging from zero (blue), if $i$ is synchronized with the mean phase, to $\pi$ (red) if $i$ is in anti-phase with $\langle \phi \rangle$).
(b) Perturbing the optimal network, by systematically removing a fraction $f$ of links, we observe a gradual degradation of synchronization. As expected, we find that for $f = 0$ we have $r = 1$, gradually decreasing as $f$ is increased.
(c) For $f \approx 0.9$, an extreme perturbation to the predicted network, synchronization is fully degraded, with individual node phases distributed within the entire range $|\Delta \phi| = 0$ (blue) to $|\Delta \phi| = \pi$ (red).}
\label{fig:Crtical_coupling}
\end{figure}
 
\noindent 
\textit{Designing networks for synchronization}. 
Our theory, up to this point, focused on the selection of $\boldsymbol{\omega}$ that will enable (\ref{eqn1}) to reach synchronization, namely, we begin with a given weighted network $A_{ij}$ and phases $\alpha_{ij}$, for which we seek the optimal set of natural frequencies $\omega_i$. Often, however, we are confronted with the opposite challenge: given a set of oscillators with natural frequencies $\boldsymbol{\omega}$, can we design a weighted network $A_{ij}$ with lags $\alpha_{ij}$ that will drive the oscillators toward synchronization? As indicated by Eq.\ (\ref{wt_freq_set2}) this reverse challenge is not as well-defined, allowing a broad degree of freedom to select the network, its link weights and the matching phase-lags, hence for a given set $\boldsymbol{\omega}$, one can construct many synchronizable networks. To examine this systematically we consider the case where the phase-lags were all set to $\alpha_{ij} = \alpha = 0.1$, leaving us with the degree of freedom to construct $A_{ij}$. Extracting the natural frequencies from a uniform distribution $\omega_i \sim \mathcal{U}(2,30)$, we constructed a weighted network that satisfies (\ref{wt_freq_set2}), by setting its weighted degrees $d_i$ (\ref{degree}) to conform with the condition (\ref{wt_freq_set}). As predicted, we find that the designed network leads to perfect synchronization $r = 1$ (Fig.\ \ref{fig:Crtical_coupling}a); perturbing this network results in gradual synchronization loss (Fig.\ \ref{fig:Crtical_coupling}b).

\begin{figure}
\includegraphics[height=4.5cm,width=8.5cm]{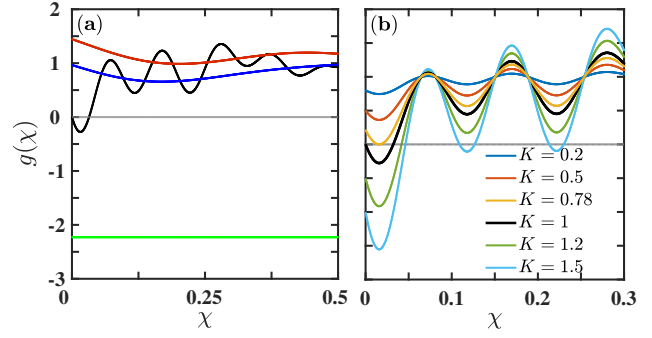}
\caption{{\bf Collective coordinate analysis of our predicted synchronization}. 
(a) $g(\chi)$ vs.\ $\chi$ as obtained from Eq.\ (\ref{gottwald4}) for the four different frequency sequences, uniform (red), normal (blue), homogeneous (green) and our predicted optimal frequency set (black). Stable synchronization is  obtained only for the optimal frequency set, as captured by the fact that $g(0) = 0$, in concordance with $g^\prime(0) < 0$. All other curves (red, blue, green) do not intersect the horizontal axis, lacking a potentially stable synchronous solution.
(b)$g(\chi)$ vs.\ $\chi$ as obtained for the optimal frequency set under different coupling strengths $K$. Stable phase locking occurs when the curve crosses $g(\chi) = 0$, observed for the first time when $K = 0.78$, the point of the onset of synchronization, as shown in Fig.\ \ref{fig:r vs k}b (yellow). }
\label{stability}
\end{figure}   

\par 
\noindent
{\it Dimension reduction analysis.} We use a {\it collective coordinate approach} \cite{Gottwald,Pinto-Saa,Brede}, to analyze the behavior of the synchronized and phase-locked solutions in the rotating frame. In this approach, the instantaneous phase of each phase-locked oscillator is approximated by 
\begin{equation}
\theta_i(t)=\chi\omega_i,
\label{gottwald1}
\end{equation}
where $ \chi = \chi(t)$ is a time dependent function. A stable phase-locked solution emerges in the network when the error
\begin{equation}
\epsilon_{i}(\chi) = \dot{\chi}\omega_i-\omega_i-\sum_{j=1}^{N} A_{ij}  \\
F(\chi(w_j-w_i)-\alpha_{ij}),
\label{gottwald2}
\end{equation}     
is minimized. Such minimization is enabled if $\epsilon_{i}({\chi})$ is orthogonal to the tangent subspace of the solution space of Eq.\ (\ref{gottwald1}), spanned by ${\frac{\partial \theta_i}{\partial \chi}}=w_i$\cite{Gottwald,Pinto-Saa}. Projecting $\epsilon_{i}(\chi)$ onto this restricted subspace and using the orthogonality condition we obtain a reduced one dimensional differential equation for $\chi(t)$
\begin{equation}
\frac{d{\bf\chi}}{dt}= g(\chi)
\label{gottwald3},
\end{equation}
where 
\begin{equation}
g(\chi)=1+\frac{1}{\sigma^2}\sum_{i=1}^{N}\omega_i\sum_{j=1}^{N} A_{ij}F(\chi(w_j-w_i)-\alpha_{ij}) 
\label{gottwald4}
\end{equation}
and $\sigma^2=\sum_{j=1}^{N}{\omega_j}^2$. Equation (\ref{gottwald3}) reaches a stable fixed point if for some choice of $\boldsymbol{\omega}$, Eq.\ (\ref{gottwald4}) satisfies $g(\chi)=0$ and $g^\prime(\chi) = \frac{d {g}}{d{\chi}}< 0$. Under these conditions we have $\chi(t) = \chi$ independent of time, which in (\ref{gottwald1}) predicts a time independent $\theta_i$, representing a phase-locked system, where all nodes oscillate at a common frequency, with their relative phases constant in time. If such fixed-point occurs for $\chi \rightarrow 0$ the phases (\ref{gottwald1}) approach a single value $\theta_i \rightarrow 0$, representing a convergence to prefect synchronization $r \rightarrow 1$. In Fig.\ \ref{stability}a we show $g(\chi)$ vs.\ $\chi$ taking our predicted optimal $\boldsymbol{\omega}$ from (\ref{wt_freq_set2}). We find that a stable fixed point ($g(\chi) = 0$, $g^{\prime}(\chi) < 0$) occurs at $\chi = 0$, precisely the predicted global synchronization obtained for under our optimal $\boldsymbol{\omega}$ (\ref{wt_freq_set2}). For the other three selections of ${\boldsymbol{\omega}}$ (uniform, normal, homogeneous), we observed that $g(\chi)$ never crosses zero, indicating that stable synchronization is indeed unattainable (Fig.\ \ref{stability}a, red, blue, green, comparing to Fig. \ref{fig:r vs k}a). 
 	
\par
To test the impact of changes to the coupling $K$, we shown, in Fig.\ \ref{stability}b, the behavior of $g(\chi)$
under the optimal ${\boldsymbol{\omega}}$, for a selection of $K$ values. 
As explained above, perfect synchrony occurs when $g(\chi) = 0$ and $g^\prime(\chi) < 0$ at precisely $\chi = 0$, conditions that are only satisfied for $K = 1$ (black), the value for which our optimal $\boldsymbol{\omega}$ was constructed. Especially important is the critical $K = 0.78$ (yellow), representing the coupling where (\ref{gottwald3}) assumes a stable fixed-point for the first time. This point represents the onset of a phase-locked solution, where $r$ begins to rise above zero. Beyond this point $r$ consistently increases as the crossing point ($g(\chi) = 0$) approaches closer to $\chi = 0$, eventually reaching perfect synchronization at $K = 1$, for which $g(0) = 0$. Indeed, in Fig.\ \ref{fig:r vs k}b we observe the onset of synchronization at precisely $K = 0.78$ (vertical yellow line), a reassuring consistency with Fig.\ \ref{stability}b.  
For larger values of $K$ ($K = 1.2, 1.5$) the system continues to exhibit a stable phase-locked solution, indicated  by the slow decline in Fig.\ \ref{fig:r vs k}b for $K > 1$ and by the consistent crossing of $g(\chi) = 0$ in Fig.\ \ref{stability}b (green, cyan).  Note, that for $K = 1.5$ the fixed-point condition is satisfied for two values of $\chi$, indicated by the two crossing points $g(\chi) = 0$, both featuring a negative slope. In our analysis, however, we only regard the first crossing point, as phase-locking is only captured by (\ref{gottwald1}) in the limit of small $\chi$.

\noindent
\textit{The onset of synchronization}. The optimal set $\boldsymbol{\omega}$ is designed for \textit{perfect} synchronization $r = 1$ at a given weight $K = K_{\rm Opt}$ (set to unity in our analyses up to this point). However, the \textit{onset} of synchronization occurs at a critical $K_c < K_{\rm Opt}$, a point where $r$ begins to rapidly ascend from the chaotic regime $r = 0$ ($K = 0.78$ in our previous example, Figs.\ \ref{fig:r vs k}b and \ref{stability}b). In Supplementary material we use mean-field analysis  \cite{Ichinomia:pre2004,Coutinho:pre87_2013,Prosenjit:pre2017} to analytically predict $K_c$ under homogeneous phase-lags $\alpha_{ij} = \alpha$ as
\begin{eqnarray}
K_c(\alpha)=\frac{2 K_{\rm Opt}^3 {\sin^{3} \alpha} {\langle k\rangle} \cos \alpha}{\pi (\Omega_c+b)^2 P(\frac{\Omega_c+b}{K_{\rm Opt}\sin(\alpha)})},
\label{cric_coup}
\end{eqnarray}
where the group angular velocity $\Omega_c$ satisfies 
\begin{eqnarray}
\pi(\frac{\Omega_c +b}{a})^2 P(\frac{\Omega_c +b}{a})\tan \alpha=\int_{q_{min}}^{\infty} \frac{q^2 P(q)}{aq-b-\Omega_c} dq,
\label{cric_omegan}
\end{eqnarray}
in which $a = K_{\rm Opt} \sin \alpha$, $b = K_{\rm Opt} \sin \alpha \frac{\sum_{i=1}^{N}q_i}{N}$, $q_{min}$ is the minimum degree over all nodes and $P(q)$ is the probability density function of the node frequencies. Equations (\ref{cric_coup}) and (\ref{cric_omegan}), our final prediction, allow us, for a given $A_{ij}$, $\alpha$ and natural frequencies $\boldsymbol{\omega}$, to express $K_c$ the critical point of transition, in which synchronization begins to emerge. Together with $K_{\rm Opt}$ of Eq.\ \ref{wt_freq_set2}) these two points fully characterize the states of the system: chaotic ($r = 0$ for $K < K_c$, phase-locked ($r > 0$) at $K \ge K_c$ and optimal ($r = 1$) at $K = K_{\rm Opt}$.
 
\begin{figure}
\includegraphics[height=4.5cm,width=8.5cm]{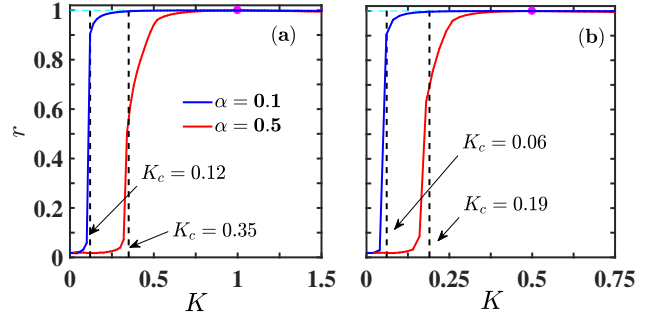}
\caption{{\bf The onset of synchronization}. We constructed a scale free network ($N = 5,000$, $\gamma = 3$, $\langle k \rangle = 30$) with homogeneous phase frustration ($\alpha_{ij} = \alpha$). 
(a) $r$ vs.\ $K$ for $\alpha = 0.1$ (blue) and $\alpha = 0.5$ (red), as obtained from Eq.\ (\ref{eqn1}). Perfect synchronization is obtained at $K_{\rm Opt} = 1$; the onset of synchronization occurs at $K_c = 0.12$ and $0.35$ (vertical dashed lines), in agreement with prediction (\ref{cric_coup}).
(b) We constructed a different selection of $\boldsymbol{\omega}$, for which perfect synchronization occurs at $K_{\rm Opt} = 0.5$. Also here, synchronization onsets at the predicted $K_c$ (vertical dashed-lines).}
\label{fig:Crtical_couplingll}
\end{figure}

To observe this we constructed a scale-free $A_{ij}$ ($N = 5,000$, $\langle k \rangle = 30$, $P(k) \sim k^{-\gamma}$, $\gamma = 3$) with homogeneous phase-lags $\alpha$. We matched this networks with the appropriate optimal frequency sets $\boldsymbol{\omega}$, such that perfect synchronization occurs at $K_{\rm Opt} = 1$. 
In Fig.\ \ref{fig:Crtical_couplingll}a we show $r$ vs.\ $K$, for 
$\alpha = 0.1$ (blue) and 
$\alpha = 0.5$ 
(red) finding that indeed, in both cases, 
$r = 1$ at the optimal 
$K = K_{\rm Opt} = 1$. 
The crucial point is that synchronization begins to appear at significantly lower values of $K$, at the critical 
$K_c$, where $r$ begins to sharply incline. We next used (\ref{cric_coup}) to calculate $K_c$ in both cases, confirming our analytical predictions, which perfectly match the observed criticality (vertical dashed lines). Repeating the same experiment, this time setting $K_{\rm Opt} = 0.5$ we observe further confirmation of our prediction (Fig.\ \ref{fig:Crtical_couplingll}b).   

Understanding the phenomena of synchronization in networks has crucial applications in fields ranging from neuronal networks to power supply. These systems are often described by highly heterogeneous weighted networks and exhibit distributed lag-times, a combination that rarely succumbs to analytical treatment \cite{Baruch-Binomial_12,Baruch-natphys,Baruch_natcommm_15,Baruch_natcommm_17}. Our analysis here has shown how to analytically construct the appropriate frequency sequence to ensure  perfect synchronization, relating the optimal frequency set to the weighted network structure and to the distributed lags $\alpha_{ij}$. It shows that the systems' weighted degree distribution plays an important role in determining the desired frequencies, where degree heterogeneity dictates a similar heterogeneity in the frequency set, a surprising result, showing that synchronization can occur by introducing diversity in $\omega_i$, rather than by increasing its homogeneity. Hence cooperative phenomena may emerge even in the presence of microscopic diversity, a consequence of the phase-lags, that is absent in the classical Kuramoto framework.    

\section{Acknowledgements}
\par The authors would like to thank  Syamal Dana for interesting comments and suggestions. P.K. acknowledges support from  DST, India under the DST-INSPIRE scheme (Code: IF140880). C.H. is supported by the CHE/PBC, Israel. This work was supported by NSF-CRISP Award Number:  1735505.

\pagebreak

\onecolumn

\textbf{\huge Supplementary of  ``Perfect synchronization in networks of phase-frustrated  oscillators"}

\section{Critical coupling strength for transition to perfect synchronization in networks with homogeneous $\alpha$ ($\alpha_{ij} = \alpha$)}
Based on the approach proposed in \cite{Prosenjit:pre2017} here we analytically derive the self-consistent equations for critical coupling strength and group angular velocity for the onset of perfect synchronization. 
For that we consider  a system of
$N$ 
coupled oscillators, whose phases $\theta_i(t)$  are driven by the  dynamic equations
\cite{Kuramoto,Sakaguchi}
 \begin{eqnarray}\label{eqn1}
\frac{d\theta_i}{dt} &=& \omega_i + \sum_{j=1}^{N} A_{ij}\sin(\theta_j - \theta_i-\alpha), ~i = 1\dots N, 
\end{eqnarray}
where $A_{ij}$ is the $ij$th element of the weighted adjacency matrix and $\alpha$ is the homogeneous phase lag parameter. We now assume that $A_{ij} = K$, if $i$th and $j$th nodes of the network are connected and otherwise $A_{ij} = 0$. In this environment of homogeneous $\alpha$ and presence of $K$, the set of optimal frequency for achieving perfect synchronization at a specific coupling strength $K = K_{opt}$ will look like (follow the Eqn.\ (8) from the main text) 
 \begin{eqnarray}
 	{{\omega}_i} = K_{opt}q_i\sin \alpha -(K_{opt} \sin \alpha \frac{\sum_{i=1}^{N}q_i}{N}) = aq_i -b, (i = 1\dots N)
 	\label{optimal_freq_at_K_{opt}_c}
 \end{eqnarray}
where $q_i=\sum_{j=1}^{N}A_{ij}/K$, $a=K_{opt} \sin \alpha$ and $b= K_{opt} \sin \alpha \frac{\sum_{i=1}^{N}q_i}{N}$.
\noindent
To quantify the order of  synchronization  we use a generalized order parameter \cite{Coutinho:pre87_2013}
\begin{eqnarray}
r(t)e^{i\psi(t)} = \frac{\sum_{j=1}^{N} q_{j}e^{i\theta_j}}{\sum_{j=1}^{N} q_{j}},
\label{op}
\end{eqnarray}
where 
$r(t)$ and $\psi(t)$  
respectively denote the order parameter and the average phase of the collective dynamics at time 
$t$. The values of the order parameter 
$r(t)$ varies in the range 
$0\leq r(t)\leq 1$. The order parameter takes the value 
$r(t) = 0$ for incoherent solution, while 
$r(t) = 1$ indicates fully synchronized state of the system.   
Following the mean-field approach proposed in \cite{ichinomia:pre_2004}, let the density of the nodes with phase 
$\theta$ at time 
$t$ for a given degree 
$q$ be given by the function 
$\rho(q,\theta,t)$, where 
\begin{eqnarray}
\int_{0}^{2\pi} \rho(q,\theta,t)d\theta =1.\label{density}
\end{eqnarray}
We assume that there is no degree correlation between the nodes of the network and therefore the probability that a randomly chosen edge is attached to a node with degree $q$ and phase $\theta$ at time $t$ can be written as 
\begin{equation}
\frac{qP(q)\rho(q,\theta,t)}{\int qP(q)dq},
\end{equation}
where $P(q)$ is the distribution of
nodes with degree $q$.
In the continuum limit, Eq.\ \ref{eqn1}  can be written as 
\begin{eqnarray}
\frac{d\theta(t)}{dt} & = & \omega  + \frac{K q}{\langle q \rangle} \int dq' \int q' P(q') \rho(q',\theta',t) \sin(\theta' -\theta -\alpha)d\theta',
\label{eqn3}
\end{eqnarray}
where $\langle q\rangle = \int qP(q)dq$ is the mean degree of the network. 
Now for the conservation of the oscillators for Eq.\ (\ref{eqn1}), the density function $\rho$ satisfies the continuity equation
\begin{equation}
\frac{\partial \rho}{\partial t} + \frac{\partial}{\partial \theta}(\rho v) = 0,\label{continuity}
\end{equation}
where $v$ is the right hand side of the Eq.\ (\ref{eqn3}). 

To measure the macroscopic behavior of the oscillators, in the thermodynamic limit ($N \rightarrow \infty$) we consider the order parameter $r$ given by \cite{ichinomia:pre_2004} 
\begin{eqnarray}
r e^{i\psi} & = & \frac{1}{\langle q\rangle} \int dq \int q P(q)\rho(q,\theta,t) e^{i\theta}d\theta,\label{eqn4}
\end{eqnarray}
where $\psi$ is the average phase of the oscillators and the value of $r$ varies in the range $0\le r \le 1$. Therefore, using (\ref{eqn4}), we rewrite the Eq.\ (\ref{eqn3}) as 
 \begin{eqnarray}
\frac{d\theta}{dt} & = & 
\omega + K q r \sin(\psi -\theta -\alpha).
\label{eqn5}
\end{eqnarray}
To derive the self-consistent equation we set the global phase $\psi(t) = \Omega t$ where $\Omega$ is the group angular velocity and introduce a new variable $\phi$ with $\phi(t)=\theta(t)-\psi(t)+\alpha$. In terms of this new variable, equation (\ref{eqn5}) can be written as 
\begin{eqnarray}
\frac{d\phi}{dt} & = & 
aq -b - \Omega - K q r \sin(\phi).\label{eqn6}
\end{eqnarray}
Then the equation of continuity (\ref{continuity}) takes the form
 \begin{eqnarray}
 \frac{\partial}{\partial t} \rho(q,\phi,t) + \frac{\partial}{\partial \phi} [v_\phi \rho(q,\phi,t)]=0, \label{eqn_cont}
 \end{eqnarray}
where $v_\phi = \frac{d\phi}{dt}$. In the steady state, we have 
$\frac{\partial}{\partial t} \rho(q,\phi,t) =0$. 

\noindent Therefore, steady state solution for the density function $\rho$ is given by
\begin{eqnarray}
 \rho(k,\phi)=\begin{cases}
\delta \left(\phi-arc\sin{\left(\frac{aq-b-\Omega}{qK r}\right)}\right), & \abs*{\frac{aq-b-\Omega}{qK r}} \leq 1 \\
\frac{A(q)}{aq-b-\Omega -qK r \sin(\phi)}, & \abs*{\frac{aq-b-\Omega}{qK r}} > 1,
\end{cases}
\end{eqnarray}
 where $\delta$ 
 is the Dirac delta function and 
 $A(q)$ 
 is the normalization constant given by 
 $A(q) = \frac{\sqrt{(aq-b-\Omega)^2 - (K r q)^2}}{2\pi}$.

The first solution corresponds to the synchronous state and second solution is due to desynchronous state. Hence the order parameter can be rewritten as
\noindent 
 \begin{eqnarray}
r = \frac{1}{\langle q\rangle} \int \bigg[\int_{q_{min}}^{\infty} q P(q) \rho(q,\phi) e^{i(\phi -\alpha)} H\bigg(1-\abs*{\frac{aq-b-\Omega}{qK r}}\bigg) dq+ \nonumber \\
 \int_{q_{min}}^{\infty}q P(q) 
 \rho(q,\phi) e^{i(\phi-\alpha)}  H\bigg(\abs*{\frac{q-\Omega}{qK r}}-1\bigg)dq \bigg] d\phi, 
 \label{r_d_l}
 \end{eqnarray}
\noindent where $H$ denotes heaviside function. Here the first part of right hand side of Eq.\ (\ref{r_d_l}) gives the contribution of locked oscillators  and the second part denotes the contribution of drift oscillators  to the order parameter $r$.

Therefore, the contribution of locked oscillators to the order parameter is
\begin{eqnarray}
r_{1} &=& \int_{q_{min}}^{\infty} \bigg[\frac{\cos \alpha}{\langle q\rangle} q P(q) \sqrt{1-\left(\frac{aq-b-\Omega}{K r q}\right)^2} + \frac{\sin \alpha}{\langle q\rangle} q P(q) \frac{aq-b-\Omega}{K r q} \bigg] \nonumber \\ 
&& \times H\left(1-\abs*{\frac{aq-b-\Omega}{qK r}}\right) dq -\int_{q_{min}}^{\infty} i\bigg[ \frac{\sin \alpha}{\langle q\rangle} q P(q) \sqrt{1-\left(\frac{aq-b-\Omega}{K r q}\right)^2}  \nonumber \\ && -\frac{\cos \alpha}{\langle q\rangle}  q P(q) \frac{aq-b-\Omega}{K r q} \bigg] H\left(1-\abs*{\frac{aq-b-\Omega}{qK r}}\right)dq,
\label{r_lock}
 \end{eqnarray}
 and that of the drift oscillators is 
 \begin{eqnarray}
r_{2}&=&\frac{(\sin \alpha +i\cos \alpha)}{\langle q\rangle}\int_{q_{min}}^{\infty} \frac{aq-b-\Omega}{K r} P(q) \left[1-\sqrt{1-\left(\frac{K r q}{aq-b-\Omega}\right)^2} \right]  \nonumber \\
&& \times H\left(\abs*{\frac{aq-b-\Omega}{qK r}}-1\right)dq.~~~~
\label{r_drift}
 \end{eqnarray}
 Hence we get $r=r_1 +r_2$, 
 where 
 $r_1$ and 
 $r_2$ are given by Eq.\ (\ref{r_lock}) and Eq.\ (\ref{r_drift}) respectively. 
 Comparing the real and imaginary parts we get
 \begin{eqnarray}
r\langle q\rangle &=&\cos \alpha\int_{q_{min}}^{\infty} q P(q) \sqrt{1-\left(\frac{aq-b-\Omega}{K r q}\right)^2} H\left(1-\abs*{\frac{aq-b-\Omega}{qK r}}\right)dq \nonumber \\
 && + \frac{\sin \alpha}{K r}(\langle q\rangle- \Omega)  -\sin \alpha \int_{q_{min}}^{\infty} \frac{aq-b-\Omega}{K rq} q P(q) \sqrt{1-\left(\frac{K r q}{aq-b-\Omega}\right)^2} \nonumber \\ 
 && \times H\left(\abs*{\frac{aq-b-\Omega}{qK r}}-1\right)dq,  
\label{real_r}
 \end{eqnarray}
and
 \begin{eqnarray}
a\langle q\rangle -b-\Omega &=&\int_{q_{min}}^{\infty} (aq-b-\Omega)P(q) \sqrt{1-\left(\frac{K r q}{aq-b-\Omega}\right)^2} H \left(\abs*{\frac{aq-b-\Omega}{qK r}}-1\right)dq \nonumber \\
 && +K r \tan \alpha \int_{q_{min}}^{\infty} q P(q) \sqrt{1-\left(\frac{aq-b-\Omega}{K r q}\right)^2} \nonumber \\
 && \times H\left(1-\abs*{\frac{aq-b-\Omega}{qK r}}\right) dq. 
\label{im_r}
 \end{eqnarray}
 From equation~(\ref{im_r}) using Taylor's series expansion we get
\begin{eqnarray}
\langle q\rangle -\Omega &=&K r \tan \alpha \int \displaylimits_{\frac{b+\Omega}{(a+K r)}}^{\frac{b+\Omega}{(a-K r)}} q P(q) \sqrt{1-\left(\frac{aq-b-\Omega}{K r k}\right)^2}  dq   \nonumber \\ &&+ \int_{q_{min}}^{\infty} (q-\Omega)P(q) \left\lbrace 1-{\frac{(K r q)^2}{2(q-\Omega)^2}}\right\rbrace dq.
 \end{eqnarray}
Taking the limit $r\rightarrow 0^+$ we can find
 \begin{eqnarray}
 \pi(\frac{\Omega_c +b}{a})^2 P(\frac{\Omega_c +b}{a})\tan \alpha=\int_{q_{min}}^{\infty} \frac{q^2 P(q)}{aq-b-\Omega_c} dq,
\label{cric_omegan}
 \end{eqnarray}
where $\Omega_c$ 
the critical group angular velocity at the onset of synchronization.
Now combining equations~(\ref{real_r}) and~(\ref{im_r}) we get 
 \begin{eqnarray}
r\langle q\rangle=\frac{1}{\cos \alpha}\int_{\frac{\Omega +b}{a-K r}}^{\frac{\Omega+b}{a-Kr}} q P(q) \sqrt{1-\left(\frac{aq-b-\Omega}{K r q}\right)^2}  dq. 
\label{cric_coup1}
 \end{eqnarray}
Substituting $\frac{aq-b-\Omega}{Kr}=y$, the equation~(\ref{cric_coup1})reduces to
\begin{eqnarray}
\langle q\rangle =\frac{K}{a\cos \alpha}\int_{\frac{-\Omega-b}{a+K r}}^{\frac{-\Omega-b}{a-K r}} (\frac{\Omega+b+K r y}{a}) P(\frac{\Omega+b+ Kr y}{a})\sqrt{1-\frac{y^2}{(\frac{\Omega+b+ Kr y}{a})^2}}  dy
\end{eqnarray}
and in the limit 
$r\rightarrow 0^+$ 
we get
 \begin{eqnarray}
K_c(\alpha)=\frac{2 a^3\langle k\rangle \cos \alpha}{\pi (\Omega_c+b)^2 P(\frac{\Omega_c+b}{a})},
\end{eqnarray}
Now substituting $a=K_{opt}\sin\alpha$ 
we have 
 \begin{eqnarray}
K_c(\alpha)=\frac{2 K_{opt}^3 {\sin^{3} \alpha}    {\langle k\rangle} \cos \alpha}{\pi (\Omega_c+b)^2 P(\frac{\Omega_c+b}{K_{opt}\sin(\alpha)})},
\end{eqnarray}
where $b= K_{opt} \sin \alpha \frac{\sum_{i=1}^{N}q_i}{N}$.

\section{Impact of distributed phase-lags ($\alpha_{ij}$) on the combined distribution of degrees ($d$) and frequencies ($\omega$)}
\begin{figure}
 	\includegraphics[height=5cm,width=18cm]{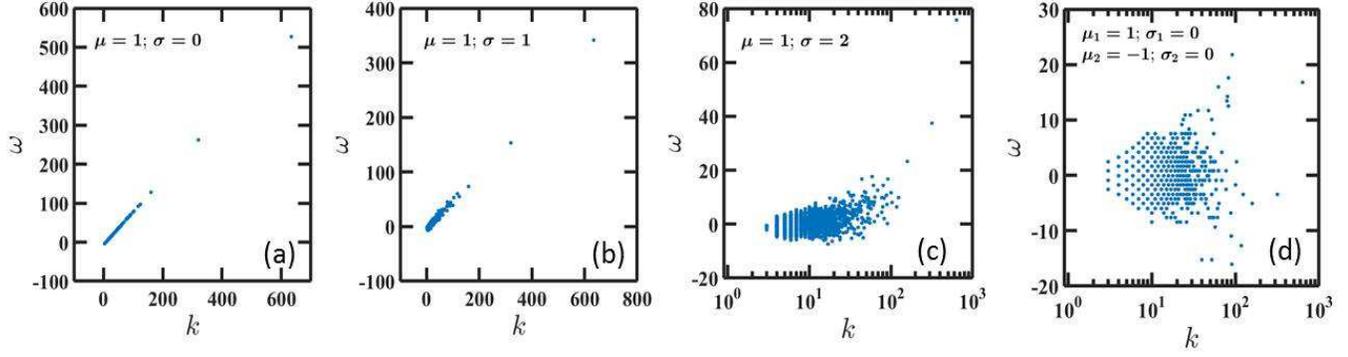}
 	\caption{{\bf The impact of phase-lags.} (a) The degree ($k$) and frequency ($\omega$) are linearly related if we set the $\alpha$ as identical ($\alpha_{ij}=\alpha=1$ as $\mu=1$ and $\sigma=0$). The linear relation is slightly broken for small deviation of $\alpha$ from $1$ ($\sigma=1$) shown in (b) where as  strong deviation from $\mu=1$ ($\sigma=2$)  and a bimodal distribution ($\mu_1=1,\sigma=0$; $\mu_1=-1,\sigma=0$)   breaks the linear relation  shown in (c) and (d).} 
 	\label{Suuplfig:alpha_imapct}
 \end{figure}

We have reconstructed the optimal frequency set ${\boldsymbol{\omega}}$ for diverse set of distributed phase-lags ($\alpha_{ij}$) using the equation (see the Eqn.\ (8) in main text)
\begin{equation}
\omega_{i} = -\sum_{j=1}^{N} A_{ij}F(-\alpha_{ij})+\frac{1}{N} \sum_{i,j=1}^{N} A_{ij}F(-\alpha_{ij}).
\label{wt_freq_set2}
\end{equation}
 We use a static scale-free network of size $N=5000$ where average degree $\langle k \rangle = 6$, $P(k) \sim k^{-\gamma}$, $\gamma = 3$). We have shown the effect of $\alpha_{ij}$ on the relation $\omega_i$ vs. $k_i$ in the Fig \ref{Suuplfig:alpha_imapct}. The optimal frequencies and the degrees are  linearly related ($\omega_i \sim k_i$) if we draw the phase lags from  $\alpha\sim \mathcal{N}(1,0)$ signifying all the phase lags are identical ($\alpha_{ij}=1$) and positive (Fig.\ \ref{Suuplfig:alpha_imapct}(a)). The degree-frequency relation is slightly disturbed when we generate the phase-lags from   $\alpha\sim \mathcal{N}(1,1)$ shown in Fig \ref{Suuplfig:alpha_imapct}(b) having less noise in the higher degree. The linearity does not hold for strong deviations ($\sigma=2$) of phase-lags from its mean value (here $\mu=1$) shown in \ref{Suuplfig:alpha_imapct}(c). In  Fig.\  \ref{Suuplfig:alpha_imapct}(d) the phase lags are  drawn from a bimoidal distribution $\alpha\sim ({\mathcal{N}(1,0), \mathcal{N}(-1,0)})$ which sets the the distribution of the frequencies around zero.

\end{document}